\documentclass[useAMS,usenatbib,usedcolumn]{mn2e}
\usepackage{rotating}
\usepackage{lscape}
\usepackage{latexsym}
\usepackage{graphicx}
\usepackage{sidecap}
\usepackage{acronym}
\usepackage{subfigure}
\usepackage{textcomp}
\usepackage{bbding}
\usepackage{epsf}
\usepackage{amsmath,amssymb}
\usepackage{float}
\usepackage{multirow}
\usepackage{flafter}
\usepackage{placeins}
\usepackage{afterpage}
\usepackage{preview}
\usepackage{here}
\usepackage{float}
\usepackage{sidecap}
\usepackage{setspace}

\newcommand{\D}{$^\circ$}
\newcommand{\NII}{[N\,{\sc ii}]}

\newcommand{\SII}{[S\,{\sc ii}]}
\newcommand{\HA}{H{$\alpha$}}
\newcommand{\OIII}{[O\,{\sc iii}]}
\newcommand{\FeII}{[Fe\,{\sc ii}]}

\newcommand{\HB}{H{$\beta$}}

\def\p0{\phantom{0}}

\newcommand{\md}{$^\prime$}

\def\cm3{cm$^{-3}$}

\def\12{$^{12}$CO}
\def\13co{$^{13}$CO}

\def\HII{H{\sc ii}}

\def\la{{\hbox{$\lesssim$}}}

\def\fmin{\hbox{$.\!\!^{\prime}$}}

\title [G213.3-0.4 - a new Galactic SNR]
{Optical detection and spectroscopic confirmation of supernova remnant G213.0-0.6  (now re-designated as G213.3-0.4)}
\author[M. Stupar \& Q.A. Parker]
{M.~Stupar,$^{1,2}$ Q. A. Parker$^{1,2}$\\
\\
$^{1}$Department of Physics \& Astronomy, Macquarie University, Sydney, NSW 2109,
Australia\\
$^{2}$Australian Astronomical Observatory, P.O. Box 296, Epping, NSW
1710, Australia}
 \date{Accepted 2010;
      Received 2010;
     in original form 2010}

\begin{document}

 \maketitle
\begin{abstract}
During a detailed search for optical counterparts of known Galactic supernova remnants (SNRs) using the Anglo Australian
Observatory/United Kingdom Schmidt Telescope (AAO/UKST) \HA\ survey of the southern Galactic plane
we have found characteristic optical \HA\ filaments and  associated emission in the area of SNR G213.0-0.6. Although this
remnant was previously detected in the radio as a non-thermal source,  we also confirm
emission at 4850~MHz in the Parkes-MIT-NRAO (PMN)  survey and at 1400~MHz  in the NRAO/VLA Sky Survey (NVSS).
There is an excellent match in morphological structure between the optical (\HA) and  radio emission.
We subsequently obtained optical spectroscopy of selected \HA\ filaments using the South African Astronomical Observatory
1.9-m telescope which confirmed shock excitation typical of supernova remnants.  Our discovery of \HA\ emission and the
positional match with several radio frequency maps led us to reassign G213.0-0.6 as G213.3-0.4 as these co-ordinates more accurately
reflect the actual centre of the SNR shell and hence the most probable place of the original supernova explosion.
Support for this new SNR ID  comes  from the fact that the X-ray source 1RXS J065049.7-003220
is situated in the centre of this new remnant and could be connected with the supernova explosion.

\end{abstract}



\section{Introduction}
Following our detection and discovery of $\sim$20 new Galactic supernova remnants (SNRs) based first on their optical emission line signatures (e.g. \citet{stu07a,stu07b,stu08})  we have also been undertaking a systematic search for optical counterparts to known SNRs  \citep[e.g.][]{stu11}. The discovery medium is  the AAO/UKST \HA\ survey of the southern Galactic plane  \citep[SHS hereafter:][]{parker05} whose high sensitivity and fine resolution has been instrumental in the uncovering of low surface brightness \HA\  emission associated with both known and newly identified SNRs. These are often seen in the form of narrow filamentary structures  which is a usual characteristic of SNRs seen in the optical. This is one of a series of papers concerning known SNRs where we report the discovery of associated optical SNR signatures. We then obtain spectroscopic follow-up where the observed emission line ratios reveal a shock excitation mechanism confirming their likely association with known remnants.
\begin{figure*}
 \begin{center}
   \subfigure{
   \label{fig1}
   \begin{minipage}[b]{0.485\textwidth}
    \includegraphics[width=245pt,height=253pt]{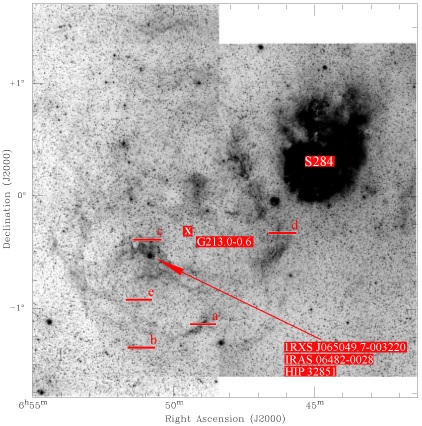}
   \end{minipage}} \hspace{-.2in}
\hspace{4mm}
   \subfigure{
   \begin{minipage}[b]{0.485\textwidth}
    \includegraphics[width=235pt,height=253pt, bb = 1 1 540 533]{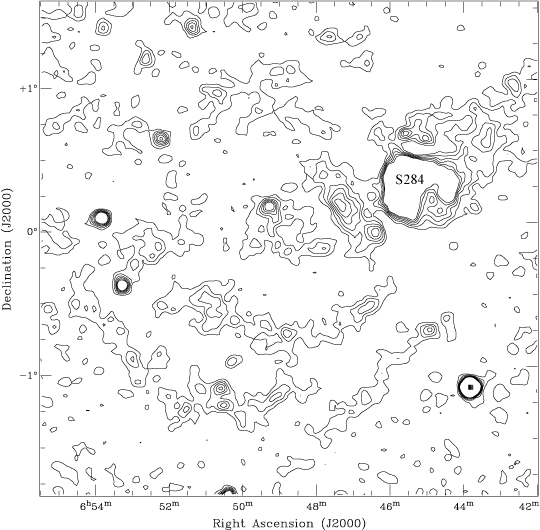}
   \end{minipage}} \\
 \caption{The left image is a high contrast high resolution 2.5\D\ \HA\ mosaic in the area of G213.0-0.6  created from the SHS \citep{parker05}. Apart from the 'burnt-out' intense emission from the compact \HII\ region S284 this high sensitivity \HA\ map reveals a fractured though coherent shell like structure to the SE of S284 with evidence of a "blowout" to the north.  The position of the previous centre of the SNR  \citep[from][]{Rei03} is indicated with an \textbf{X} and is clearly seen to be off-centre from the optical shell. Towards the central area of the shell is found the X-ray source 1RXS J065049.7-003220 and IR source IRAS 06482-0028 as well as the double-star system HIP 32851 (according to SIMBAD). Red horizontal bars mark the relative positions of the spectrograph slits for the follow-up optical spectroscopy. Their accurate locations with respect to individual optical filaments are shown in Fig.~\ref{fig5}. \HII\ region S284 is shown much brighter on this image  due to the fact that we  wanted to enhance the visibility of  optical filaments (Fig.~\ref{fig2} gives better definition of S284). The right image shows the  PMN 4850~MHz  radio contours of the same region  with flux levels between 0.01 and 0.12~Jy beam$^{-1}$. Comparison of these images shows an excellent optical/radio match. In addition, the first of two  blowout on the north  also has its  counterpart at 4850~MHz, where radio flux contours firmly follow the optical signal. Based on these optical and radio images it is clear that this SNR should be reassigned as G213.3-0.4 which better corresponds to the centre of both the optical and radio shell structures. }
 \label{fig1}
 \end{center}
\end{figure*}

The first radio investigations in the area around \mbox{G213.0-0.6} were reported in \citet{Bons79} where at 408~MHz (bandwidth 2~MHz; resolution 4\fmin2$\times$100\md )  \citet{Fan74}, using the Bologna single 32-m dish radio telescope, they made pointed observations towards G211.7-1.1, in the area of  \HII\ region S284\footnote{Actually, the first radio classification of S284 as an \HII\  region comes from \citet{Fel72} using the NRAO 300-foot telescope with resolution of 10\md at 1400~MHz.}. Using their data  and additional data for extended objects from the Bologna Catalogue  \citep{Fel78}, \citet{Bons79} realised that S284 is also surrounded by non-thermal radio emission. They then decomposed this \HII\  region  (which has an almost flat spectral index)
from the non-thermal component and derived a spectral index  $\alpha=-0.5$ for this section. This is typical of a SNR. They also
concluded (perhaps erroneously) that there is a physical association between the thermal and non-thermal emission components.

In reality however, G213.0-0.6 is situated somewhat east of S284 (where most of the faint emission is found) with an extension of approximately 160$^{\prime}\times140^{\prime}$. This can best be seen in the 863~MHz (using the 100-m Effelsberg radio telescope with resolution of some 14\fmin5) maps of
\citet{Rei03}. These maps also show that a small part of this emission appears to go around the S284 \HII\ region, in
agreement  with \citet{Bons79}.  \citet{Rei03} also mentioned that this object was also previously seen in
the Effelsberg  Galactic surveys at 21 and 11~cm, but  there are difficulties  with estimating  radio spectral
indices from this survey. Using their observations and convolved with 2.695~GHz data from the Efelsberg's survey
they estimated a spectral index \mbox{$\alpha=-0.4\pm0.15$} (applying  the TT-plot technique). This is in general agreement with the findings of \citet{Bons79},  again confirming the likely SNR nature of this more extended emission. They also found extended infrared emission north and south of S284 in the IRAS 60 and 100~$\mu$m maps of the region and partially overlapping G213.0-0.6.

On the basis of their observations, \citet{Rei03} suggested that it is an old supernova remnant, partially
shell-like in structure and extremely low in radio surface brightness. They suggest that if G213.0-0.6
is directly interacting with the S284~\HII\ region \citep[note an interaction of the thermal and non-thermal emission components is also supported by][]{Bons79} the distance to this object could be 2.4~kpc and therefore have a size of {\thinspace\mbox{110$\times$98~pc}}.
This would classify this SNR among the largest in the Galaxy.

The objective of this work is to present the existing radio observations of G213.0-0.6 from both the 4850~MHz  $\sim$5~arcmin resolution Parkes-MIT-NRAO
(PMN)\footnote{The Parkes-MIT-NRAO (PMN) radio survey was done with three different radio telescopes: the Parkes 64-m dish and the Green Bank 43 and 91-m dishes.}  survey \citep[see][]{cgw93}  and the 1400~MHz   of 45~arcseconds data from the NRAO/VLA Sky Survey (NVSS), see \citet{Con98}. This is done
in the context of the new \HA\ detection  of the SNR from the SHS \citep{parker05} so that their inter-relationship is clear. The optical morphological structure of G213.0-0.6 is shown to be very similar to the extant radio morphology seen with both PMN  and NVSS. Note G213.0-0.6 is also detected in the low-resolution but high sensitivity Southern \HA\ Sky Survey Atlas (SHASSA hereafter) of \citet{gaus01}  also in a coherent shell form (see Fig.4). We also present our follow-up spectral observations of selected \HA\ filaments. The resultant emission line spectra and diagnostic emission line ratios confirm the presence of shock-ionisation which is typical for SNRs.

\section{Observations}
\subsection{Optical and radio morphology}

During systematic searches for optical counterparts to known (mostly from radio observations)  Galactic SNRs in the SHS
(e.g. see \citet{stu07a} or \citet{stu07b}; \citet{stu09}) we noticed  a large shell-like structure of optical filaments and emission clouds some
{\thinspace\mbox{2\D$\times$3\D}} in size on the south-east side
of \HII\ region S284.  We associated these with the known SNR G213.0-0.6. The SHS \HA\ image of G213.0-0.6 is shown in Fig.~\ref{fig1}
where the overall impression is that of an almost complete optical shell consisting of a series of fractured but coherent emission filaments apparently centred on the X-ray source 1RXS J065049.7-003220  and IRAS source  06482-0028 (from SIMBAD). Furthermore, the brightest \HA\
emission occurs in this central area as well as on the north-west side where the shell is apparently closer to S284.
Fig.~\ref{fig1} also shows a remarkable extension to the north of the main optical shell in the form of  a set of parallel \textquotedblleft blow-outs\textquotedblright ~about 30~arcminutes apart with the second, almost one degree from
the northern boundary of the shell.  These blow-outs are not common but have sometimes been seen in optical SNRs when the emission is mostly fragmented (unless in the case of a young remnant) inside the radio borders in the form of  optical filaments and emission clouds. Typical examples are G189.1+3.0 \citep[or IC 443; see][]{Fes84} and G243.9+9.6 \citep{stu07b}. Occasionally, blow-outs  can also be seen in the radio morphologies at different frequencies \citep[an example is G301.4-1.0 in][]{MOST}.  Several faint emission
clouds are also present on the south-west side of S284 although  any connection of this emission with G213.0-0.6 or S284
is dubious. After discovery of this optical shell in the SHS  we examined the equivalent area in the low resolution but high sensitivity SHASSA atlas of \citet{gaus01}. The SHASSA survey resolution is only 48~arcsecond/pixel but this has the effect of blending some of the fine \HA\ filaments together making the \HA\ shell like appearance of this SNR even more striking. Indeed at the lowest \HA\ contours the emission forms almost a complete optical shell apart from at the location of the blow-out to the north. This can be clearly seen in Fig.~\ref{SHASSA} and indicates that the optical counterpart to this SNR could easily have been discovered in SHASSA.

\begin{figure}
\center
\includegraphics[width=8.4cm,height=8.3cm, bb = 1 1 537 538, clip=,]{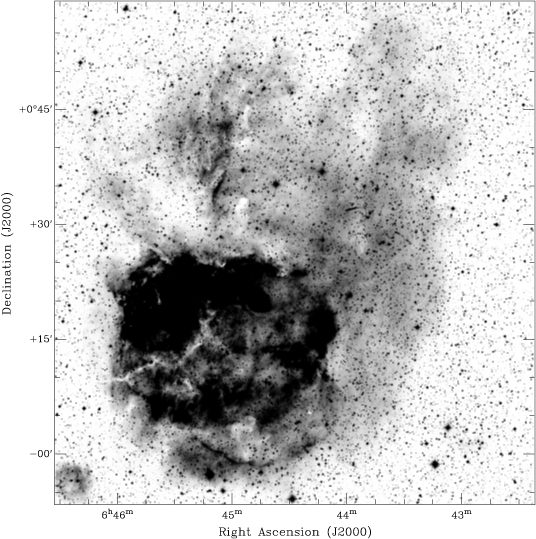}
\caption {High resolution,  lower contrast image of the S284 \HII\ region from the SHS.
It is clear that the  diffuse emission north from the bright central section and the bright central region itself has
a typical \HII\ region morphology  in terms of internal dust-lanes and turbulent structures and that this is highly unlikely to form part of an SNR. }
\label{fig2}
\end{figure}

\begin{figure}
\center
\includegraphics[width=8.4cm,height=8.3cm, bb = 1 1 533 543, clip=,]{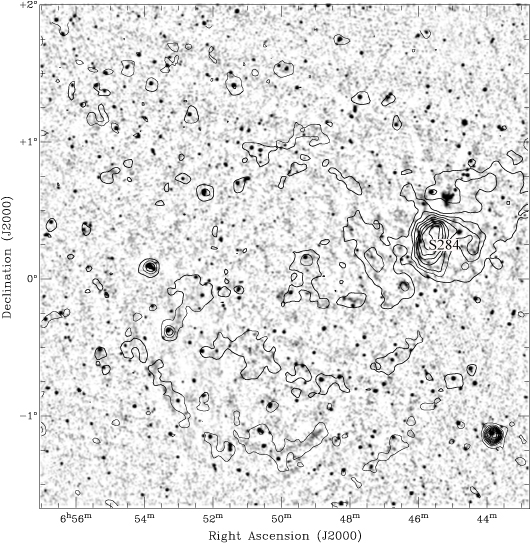}
\caption {The area around G213.0-0.6  (G213.3-0.4) from the higher resolution 1400~MHz NVSS radio survey overlaid  with  contours from the
lower resolution 4850~MHz PMN data used to mark the position of the very low flux detection of this remnant at 1400~MHz.
In the NVSS the southern and central area  of G213.0-0.6 is clearly seen to have
an excellent match to both the 4850~MHz PMN data and the \HA\ optical emission but with a more  incomplete  match to the northern blow-out.
The brightest point like sources at 1400~MHz have flux  around  0.3  Jy beam$^{-1}$ while the brightest filaments bounded with 4850~MHz contours have very low flux of some  0.002 Jy beam$^{-1}$. }
\label{fig3}
\end{figure}

We also show a deep, high resolution \HA\ image of the S284 \HII\ region (see Fig.~\ref{fig2}) where the internal dust lanes and turbulent nature of the emission is shown clearly for the first time and is shown to be completely typical of \HII\ regions.
The previous radio observations of the SNR also overlapped this object  (most probably due to low angular resolution at the given radio
frequencies) and with other diffuse nebular emission to the north \citep[especially in the radio observations of][where  the north-south coverage was 110~arcminutes]{Bons79}. The high quality SHS \HA\ image reveals that S284 actually consists of an approximately circular bright component
some 30-arcmin across  centred at J2000  \mbox{RA= 06${^h}$~45${^m}$~08.7${^s}$ $\delta$= +00\D\ 14'~04''}.  There is also a  fainter, extensive, but coherent emission component extending on 3 sides of the main \HII\ region approximately 7~arcminutes to the south, about 10~arcminutes to the west and about 30~arcminutes to the north-west. The older Palomar Observatory Sky Survey (POSS-I) broad-band red image of this area shows the central bright part with much less detail. The surrounding more diffuse component is barely detected as a few isolated patches above the noise  and with an overall  diameter of some 60~arcminutes. The more recent UKST R-band survey of the region reveals more detail but it is the SHS data that finally reveals the clear \HII\ region nature of the main component and associated outer region around it. It is also clear from Fig.~\ref{fig2} that the S284 complex is morphologically and spatially distinct from the extended filamentary emission that we associate with the SNR.  Hence  this  emission cannot be accepted as a part of the SNR as has been suggested in the previous radio observations. The question of any possible interaction between the SNR and \HII\ region remains.
\begin{figure}
\center
\includegraphics[width=8.4cm,height=8.3cm]{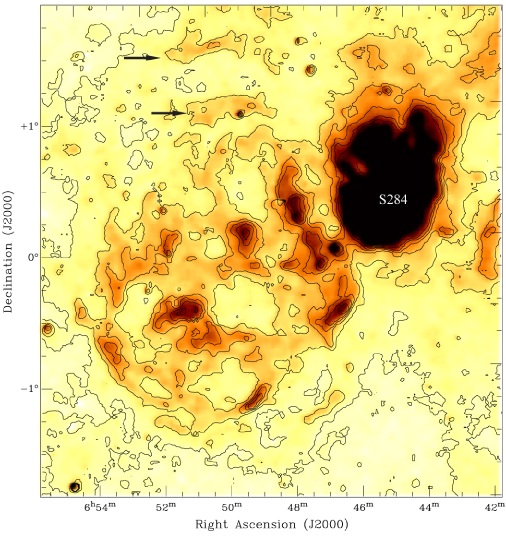}
\caption {The G213.0-0.6 area as registered by SHASSA. The SHASSA resolution is low (48~arcsecond/pixel) but we can clearly see a complete optical shell at the lowest sensitivities which is clearly separated from the intense S284 \HII\ region over 1\D\ to the north-west.  The same structure evident in the SHS image shown  in Fig.~\ref{fig1} can be clearly seen, including the  parallel emission bars of the northern blow-out 30~arcminutes apart.  Two arrows show the locations of these possible blow-outs which can be compared with previous radio images. The main SNR optical counterpart could also have been easily discovered in this survey. The SHASSA contours are from 150 to 300 decirayleighs (dR) where 1~dR = 0.1~R.}   \label{SHASSA}
\end{figure}

\begin{table}
\centering \setlength{\tabcolsep}{1.6mm}
\scriptsize{
\caption{South African Astronomical Observatory 1.9-m telescope
log for spectral observations of G213.0-0.6 in December 2008.  For all observations
a low dispersion grating of 300 lines~mm$^{-1}$ was used so as to cover all the main optical emission lines from 4500-6800\AA\ apart from \OIII\ 3727\AA.}
\setlength{\tabcolsep}{1pt}
\begin{tabular}{ccccccc}
\hline\noalign{\smallskip}
Tel.&Figs. \ref{fig5} \& 6&Date  & Exposure &Spectral&Slit&Slit  \\
& reference&& (sec)&~range (\AA)&RA&Dec.\\
\hline\noalign{\smallskip}
&&&&&$~~~~^h$~~$^m$~$~^s$&~~~~~\D~~~'~~~''\\[-3pt]
1.9-m &a)&17/12/2008&1200&4500--6900&06~49~09&--01~10~30\\
1.9-m &b)&18/12/2008&1200&4500--6900&06~51~22&--01~21~17\\
1.9-m &c)&18/12/2008&900&4500--6900&06~51~33&--00~23~03\\
1.9-m &d)&19/12/2008&900&4500--6900&06~46~16&--00~20~07\\
1.9-m &e)&21/12/2008&1200&4500--6900&06~51~13&--00~57~03\\
\hline
\end{tabular}
}
\vspace{-5pt}
\label{table1}
\end{table}

Note that our visual searches of the SHS uncovered this candidate SNR which we then attempted to match against known remnants. Our search was conducted in this fashion so as to be an unbiased trawl for SNR candidates from the SHS based purely on their optical emission signatures rather than directly searching the SHS for matches with known remnants. During subsequent searching of the PMN 4850~MHz radio survey data for a radio counterpart  to this optical candidate we discovered
that at this frequency the observed radio flux actually completely followed the  \HA\ emission. Note that this is not always the case with evolved and
old  remnants but is  common in  young SNRs. There was also an excellent match with the optically seen blow-outs to the north (see right
panel in Fig.~\ref{fig1}).
Furthermore, from the PMN radio image, it is clear that S284 and G213.0-0.6 are clearly separate objects though a possible
interaction between them is not completely ruled  out (see later discussion). We note that G213.0-0.6 is in the form of a typical shell remnant with a clearly defined arc  to the south-east  and another one on the north-west which partially  deviate from spherical curvature.
The interior region, where the SN explosion occurred, contains a rich chain of  arcuate
filaments that can be seen in the optical \HA\ imagery with a slight difference compared with the radio data. This is partially due to the median resolution of the PMN survey ($\sim$5~arcmin) though exact positional correspondence between individual optical and radio filaments is not expected (e.g. see Cram, Green \& Bock, 1998).
Note that the radio image does not reveal any point-like sources at this frequency at the proposed new centre of the remnant  nor at, or around, the previous G213.3-0.4 position. However, X-ray and infrared sources do exist close to the remnants new centre position as previously mentioned.
Taking into account the large size  of this remnant ($\sim$2\D\  plus the northern blow-outs) and the fragmented nature of the clouds and filaments in the higher resolution optical and associated radio data, we form the  overall  conclusion that this is an old remnant well on the way to dissolving into and mixing with the ISM.

This statement can be confirmed with observations of this object at 1400~MHz (see Fig.~\ref{fig3}) from the NVSS radio survey where the apparent size and shape are almost identical to both the \HA\ and PMN radio data at 4850~MHz. This match can be seen in  Fig.~\ref{fig3} where very low radio emission (compared with PMN) is detected at this frequency.   The NVSS signal appears strongest across the central and southern parts of the SNR and partially over the  northern blow-out.

\begin{table*}
\centering 
\scriptsize{
\caption{Ratio of the observed line intensities$^\dagger$ taking \HA=100 from all the observed optical spectra listed in Table~1, including
the ratio of the \NII\ and \SII\ lines against \HA\ and the ratio of \SII\ 6717/6731\AA\ lines which can provide an estimate of electron density when not in the high or low density limits.}
\setlength{\tabcolsep}{1pt}
\begin{tabular}{@{\extracolsep{8pt}}ccccccccccccc}
\hline\noalign{\smallskip}
Date&Figs. \ref{fig5} \& 6&\HB\  & \NII &\HA&\NII&\SII&\SII&\NII/\HA&\SII/\HA&\SII & electron density  \\
& reference&& 6548\AA&&6583\AA &6717\AA &6731\AA &&&6717/6731\AA &cm$^{3}$ \\
\hline\noalign{\smallskip}
17/12/2008&a)&69&28&100$^a$&48&37&27&0.76&0.6&1.4 &$\sim$10$^{2}$\\
18/12/2008&b)&51&14&100$^b$&45&49&29&0.59&0.8&LDL &--\\
18/12/2008&c)&20&15&100$^c$&35&32&20&0.50&0.5&LDL&--\\
19/12/2008&d)&--&19&100$^d$&39&72&39&0.58&1.1& LDL &--\\
21/12/2008&e)&--&27&100$^e$&61&69&28&0.88&1.0& LDL &--\\
\hline
\end{tabular}
}
\vspace{-5pt}
\begin{flushleft}

\hspace{0.9cm} \tiny The \HA\ flux:  $^a$=1.70$\times$10$^{-14}$; $^b$=2.02$\times$10$^{-14}$; $^c$=4.47$\times$10$^{-14}$;
 $^d$=3.69$\times$10$^{-15}$; $^e$=2.96$\times$10$^{-14}$ in units of erg cm$^2$ s$^{-1}$ \AA$^{-1}$\\
\hspace{0.9cm}\tiny $^\dagger$Note the rms wavelength dispersion error from the arc calibrations was 0.08\AA while the relative percentage error in flux determination from \\
\hspace{1.1cm}\tiny calibration using brightest lines was estimated as $\sim$12\%.\\

\end{flushleft}
\label{table1}
\end{table*}

\subsection{Follow-up spectral observations}

In December 2008 we undertook a series of  5 widely separated optical spectral observations across  an area of  $\sim$2\D\  of G213.0-0.6 using a CCD spectrograph attached to the 1.9-m Radcliffe telescope of South African Astronomical Observatory (SAAO),  see Table 1. The slit width was set at
$\sim$1.5~arcseconds and observations were flux calibrated using photometric standard stars LTT1020 and LTT2415. To facilitate spectral classification and to evaluate the presence of possible shock signatures which is particularly important for categorization of SNRs,  we used a low dispersion grating of 300 lines~mm$^{-1}$ which covers a range between 3500 and 6900\AA.

\begin{figure*}
     \begin{minipage}{84mm}\centering
      \includegraphics[width=5.1cm,height=4.1cm, bb = 1 1 541 494, clip=,]{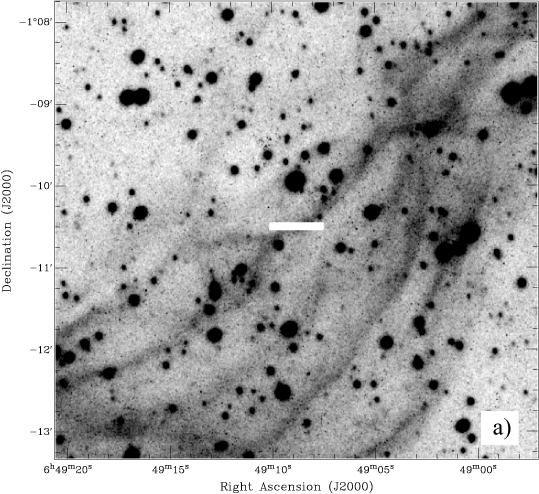}
      \includegraphics[width=5.1cm,height=4.1cm, bb = 1 1 575 518, clip=,]{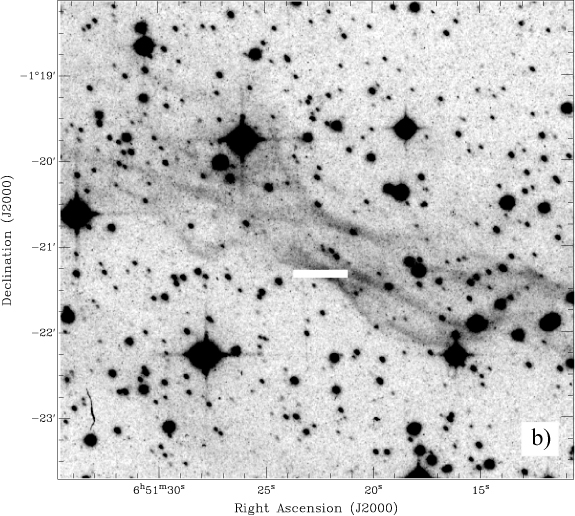}
       \includegraphics[width=5.1cm,height=4.1cm, bb = 1 1 578 538, clip=,]{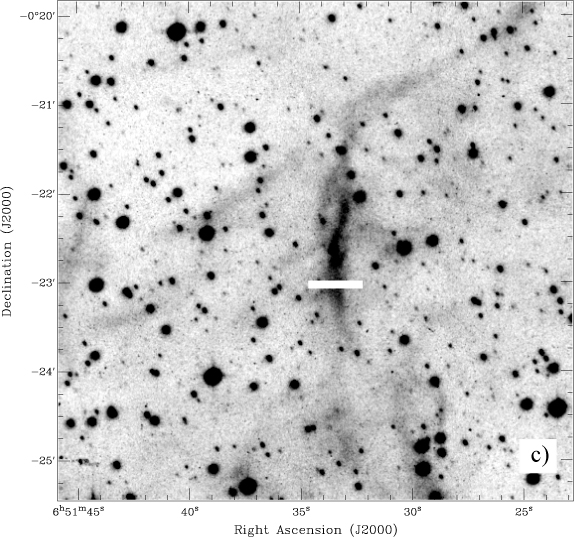}
          \includegraphics[width=5.1cm,height=4.1cm, bb = 1 1 575 538, clip=,]{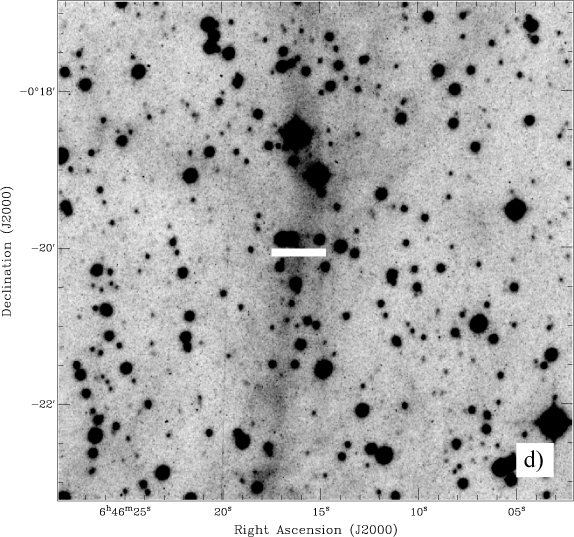}
        \includegraphics[width=5.1cm,height=4.1cm, bb = 1 1 578 538, clip=,]{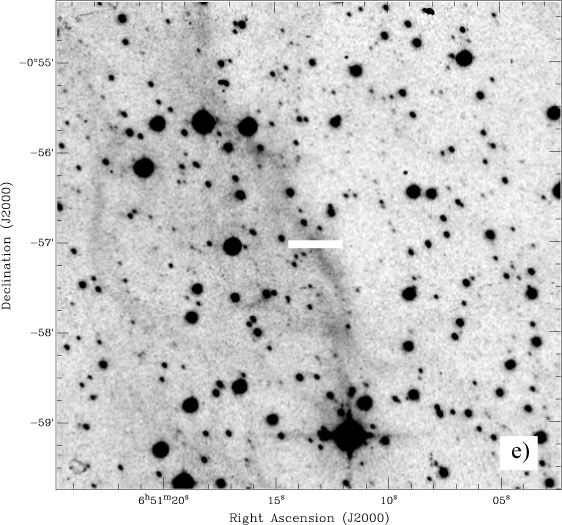}
     \caption{Extracted 5~arcminute \HA\  images from the full 0.67~arcsecond/pixel resolution SHS survey showing precise slit positions for
     all five spectral observations listed in Table~1. Filaments where selected where the  \HA\ emission appeared most prominent.
However,   due to non-photometric nights and extinction, only the strongest spectral lines in the red were evident (see also Fig.~6).}
      \label{fig5}
       \end{minipage}
          \hspace{21pt}
           \begin{minipage}{84mm}\centering
    \includegraphics[width=5.7cm,height=3.5cm, bb = 1 1 422 272, clip=,]{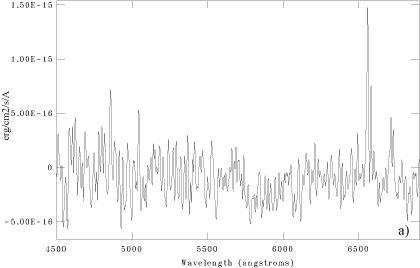}
       \includegraphics[width=5.7cm,height=3.5cm, bb = 1 1 422 270, clip=,]{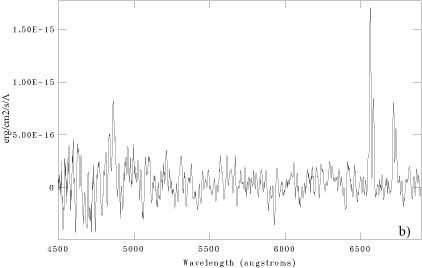}
     \includegraphics[width=5.7cm,height=3.5cm, bb = 1 1 422 270, clip=,]{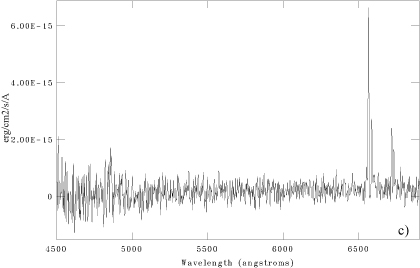}
         \includegraphics[width=5.7cm,height=3.5cm, bb = 1 1 422 270, clip=,]{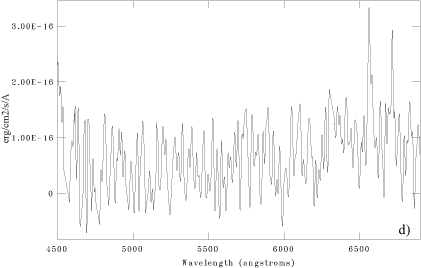}
           \includegraphics[width=5.7cm,height=3.5cm, bb = 1 1 422 270, clip=,]{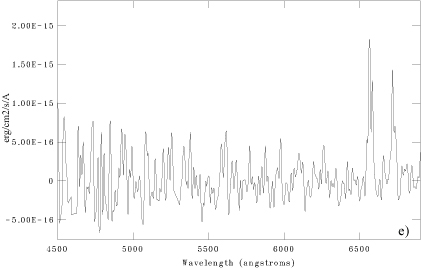}

\caption{Nominally flux calibrated 1~D spectra between 4500 and 6900\AA\  acquired as listed in Table~1 and from the slit positions marked on Fig.~\ref{fig5}. Every reference letter on the 1~D spectra corresponds to the same letter as given for the corresponding \HA\  image in Fig.~\ref{fig5}.  Every spectrum exhibits a strong \HA\ line and except for filaments labeled  \textbf{d)} and \textbf{e)} have weak, but clearly detected \HB\ but no obvious \OIII. All spectra also exhibit the \NII\  forbidden lines at 6548/6583\AA\  and especially the \SII\ lines  at  6717/6731\AA\  which are all strong relative to \HA\ and completely atypical if they originated from a \HII\ region. Indeed their ratio with \HA\ is never less than 0.5 and reveals a strong shock signature  typical of supernova remnants. Even in the case of the spectrum marked \textbf{d)} (taken on December 19, 2008), these lines are still seen even though  the spectrum is extremely noisy. Interestingly the top spectrum marked  \textbf{a)} exhibits possible weak  \FeII\ lines at 5039/5043\AA, not common in SNRs but perviously noticed in the Cygnus Loop \citep[e.g.][]{Fes96}. }

      \label{fig6}
    \end{minipage}
    \end{figure*}

Unfortunately, most observations were taken on non-photometric nights so the flux calibration is indicative only and Balmer decrement extinctions were not estimated  due to concerns of reliability (see Table~2). Moreover, in the blue  the CCD on the 1.9-m SAAO spectrograph has low response and when combined with the non-photometric nights, pushed us to avoid the bluest 1000\AA\ of the spectrum between 3500 to 4500\AA\ due to the low S/N. It is possible that some lines in the blue were lost as a result of the poor blue performance and/or extinction. Fortunately, the red part of the spectrum is well covered and is where the key diagnostic lines used to help classify objects as SNRs are located  \citep[e.g.][]{Fes85}. Furthermore, the line ratios used are based on closely separated lines which means that poor flux calibration is not a serious issue. The slit positions were carefully chosen to sample the brightest filaments seen in Fig.~\ref{fig5}, but they are still faint and when coupled with the poor weather and modest blue response led to generally noisy spectra (see Fig.~6) but still of sufficient quality to allow essential classification.

\subsection{Infrared observations}
The area around G213.0-0.6  was  reported in the infrared for the first time by \citet{Arendt89} in the survey of Galactic supernova remnants using IRAS data. \citet{Arendt89} also noted S284 as the infrared source G211.7-1.1 and properly concluded that it is an \HII\ region. However,  it is not clear if these observations also included G213.0-0.6. \citet{Sak92} continued the search for IRAS supernova remnants,  detecting the source G211.7-1.1 (or \HII\ region S284) and classifying it (wrongly)  as an SNR. The wide area around G213.0-0.6 where the SNR is actually situated is not mentioned in \citet{Sak92}  although  they  compared their infrared detection (seen in all IRAS bands) with radio observations of this  region from \citet{Fel72}.  Unfortunately,  the \citet{Fel72} radio observations were of low spatial resolution  ($\sim$10~arcminutes) so it is  likely that they could not clearly distinguish between the \HII\ region and SNR in the area.  With the better resolution NVSS data (at the same frequency of \citet{Fel72}  observations) of  G213.0-0.6  can be noticed  over the noise (see Fig.~\ref{fig3}), so our statement  that  \citet{Fel72} did not notice the remnant due to their low spatial resolution  is reasonable as is fact that this region contains  two distinct sources, SNR G213.0-0.6 and \HII\ region S284.

\begin{figure*}
 \begin{center}
   \subfigure{
   \label{fig6}
   \begin{minipage}[b]{0.485\textwidth}
\includegraphics[width=240pt,height=210pt, bb = 0 0 572 460 clip,]{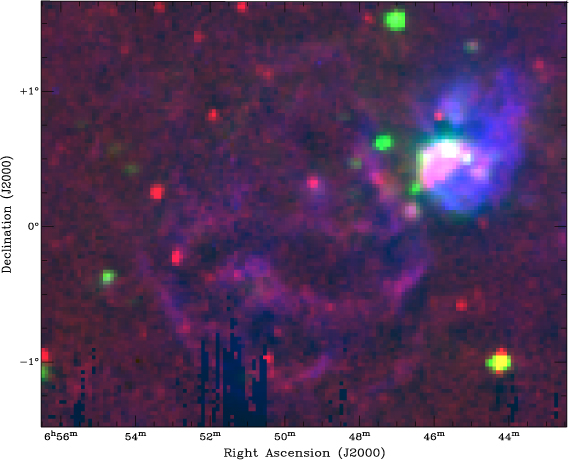}
   \end{minipage}} \hspace{-.2in}
\hspace{5mm}
   \subfigure{
   \begin{minipage}[b]{0.485\textwidth}
    \includegraphics[width=240pt,height=210pt, bb = 0 0 474 440 clip,]{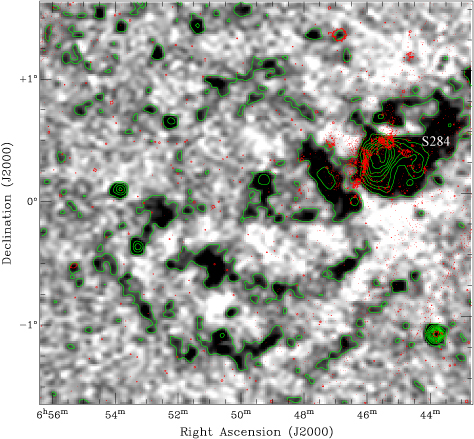}
   \end{minipage}} \\
 \caption{Left: a colour multi-wavelength montage of the IRAS (e.g IRIS) 60~$\mu$m image of G213.0-0.6 (green) with the PMN radio data (red) and the smoothed SHASAA optical emission (blue).  Only the adjacent S284 \HII\  region has any significant IRAS signature. However, it  is clear that in infrared SNR G213.0-0.6 is separated from the thermal S284  \HII\ region   opposite to conclusion of \citet{Sak92}. Right:  the area around G213.0-0.6 and S284 in  the PMN  radio survey with border contours (to show the size) overlaid in green with MSX  emission at 8.3~$\mu$m shown in red. The  SNR G213.0-0.6 is not detected at
 8.3~$\mu$m as expected for SNRs at this wavelength  (see  Discussion). The  \HII\ region S284 is a thermal source and is detected and partially spreading, but not completely matching or following the optical/radio emission  over the north-west part of G213.0-0.6.  }
 \label{fig7}
 \end{center}
\end{figure*}

To get better definition of the existing infrared data around  G213.0-0.6 we used the 'Improved Reprocessing of the IRAS Survey' data  (or IRIS)\footnote{see http://irsa.ipac.caltech.edu} at 25,60 \&100~$\mu$m (resolution from 30~arcsec to  2~arcmin). A large 8~deg  region centred on the SNR was examined which revealed no clear connection between the clearly related optical and radio data and that from the IRIS infrared observations.
Fig.~\ref{fig7} shows a colour, multi-wavelength montage of  a 4 degree region around G213.0-0.6.  The PMN 4850~MHz data is red, the IRAS 60~$\mu$m data is green while the smoothed \HA\  from SHASSA is blue. The SNR is clearly well detected in the PMN and SHASSA data while only the \HII\ region S284 is seen clearly in the IRIS 60~$\mu$m data. Furthermore,  the \HII\ region itself seems clearly separated from G213.0-0.6  calling into doubt any direct connection or interaction as suggested by  \citet{Sak92}. This throws into consideration doubt  making use of the distance to the \HII\ region to infer a physical size for the SNR (see Discussion).

Examination of the mid-infrared MSX\footnote{see http://irsa.ipac.caltech.edu/Missions/msx.html} (Midcourse Space Experiment) survey data at
8.3~$\mu$m in the area around SNR G213.0-0.6 did not show any association of 8.3~$\mu$m emission with optical or radio emission nor with the other three MSX wavelengths. This confirms  our expectation that non-thermal emission from this remnant  could barely be detected  at 8.3~$\mu$m  which is consistent  with \citet{coh01}. \citet{coh01} also showed that thermal emission from \HII\ regions must be detected by MSX at 8.3~$\mu$m, as is clearly seen in the right panel of Fig.~\ref{fig7} where red contours overlaid on the PMN 4850~MHz image mark the emission from S284. Inspection of recently published preliminary data from the  Wide-field Infrared Survey Explorer (WISE) space mission \citep{wright10} also showed no clear detection of G213.0-0.6 in any of the four WISE bands.

\section{Discussion}

The low resolution radio observations of  \citet{Bons79} and \citet{Rei03} were the first to identify G213.0-0.6 as a likely SNR. This was based on their conclusion that this object has non-thermal emission with a spectral index  $\sim-$0.5 which is typical for shell remnants. From the newly compiled radio (and new optical) images presented here it is clear that no radio emission from this remnant  extends around S284 though S284 is itself a strong radio source with a flat thermal spectrum.  This observation is in contradiction with the earlier observations noted above which suffered from low resolution and may explain the difference. The flux estimates should be re-evaluated with  further radio observations though it is anticipated that any flux correction will be small and that the observed radio spectral index will remain negative and still inside the acceptable values found for shell supernova remnants.

Table~2 presents the most prominent observed spectral line intensities for each slit position in terms of  \HA=100, as well as the ratio of \NII/\HA, \SII/\HA\ and \SII\ 6717/6731\AA. It is a commonly  adopted diagnostic  \citep[see][]{Fes85} that  an \SII/\HA\ ratio $>$0.5 can separate SNRs from  planetary nebula and \HII\ regions. All SNR spectra provide ratios well inside this criteria with \SII/\HA\  between 0.5 and $\sim$1. The observed ratios of \NII/\HA\ is a little lower compared with many other optical SNR \citep[][and references therein]{green2009}  but still acceptable, although this is somewhat dependent  on the nitrogen abundance in the Galaxy.\footnote{Conversely, there are cases where \NII\ is extremely strong in some  SNRs e.g. see \citet{stu09}.}  The large angular size and morphological structure likely indicate an old remnant and this is also supported by the observed ratio of the sulphur lines 6717/6731\AA\ which are generally in the low density limit of $\sim$~1.45 \citep{Ost06}. The exception is  for the spectrum from slit position a)  from December 17, 2008 where we can estimate Ne$\sim$10$^{2}$~cm$^{3}$. In young remnants this \SII\ ratio  goes well bellow 1 due to the higher electron densities present.  Confirmation that we have a case of an old supernova remnants comes from the fact that in some young remnants with relatively high velocities (order of $\geq$200~kms$^{-1}$) Balmer dominated shock(s) can be noticed (they are "non-radiative") without any \SII\ lines, what is certainly not the case here.

The point argued by \citet{Bons79} and \citet{Rei03} that G213.0-0.6 is most probably interacting with the adjacent, prominent \HII\ region S284  and therefore that the SNR distance can likewise be estimated to be 2.4~kpc is not bourns out by this study. \citet{Rei03} mentioned that this association follows from an apparent interaction which can be explained  if we take into account their 14~arcminute radio telescope resolution. From the high resolution SHS \HA\ imaging,  comparison with the \citet{Rei03} radio morphology image and our PMN $\sim$5~arcmin resolution  image from Fig.~\ref{fig1} and our new combined optical and radio colour image in Fig.7 it is clear there is no obvious interaction.  However, there is a clear  match between the radio and optical morphology of the SNR. Although we do not know the precise spatial distribution of the radio (or optical) filaments, it is clear from the PMN image  that  on all sides (except to the north where the blowout is present) that the shell is  effectively coherent,  including to the north-west where S284 is situated.
We might also suspect that the radio or optical shell would be fractured on the side toward S284  if any interaction exists (which is not the case) and that the low resolution of the previous radio observations would not show the fine detail we have seen from the PMN, NVSS and high resolution SHS observations.

Let us now consider the mid-infrared MSX 8.3~$\mu$m emission in the vicinity of G213.0-0.6 which is shown in the right panel of Fig.~\ref{fig7}. The MSX data can reveal emission from molecular gas and emission from warm and cool dust grains. The 8.3~$\mu$m  band is one of six MSX bands with a resolution of 18.3~arcsec  \citep[see details in][]{price01}. Besides this thermal emission, the MSX 8.3~$\mu$m band is also sensitive to fluorescent emission from polycyclic aromatic hydrocarbons (or PAHs). \citet{coh01} compared images from MSX and from radio continuum images from the 408~MHz from Molonglo Observatory Synthesis Telescope (MOST) and concluded that classical \HII\ regions are registered in both MSX bands (where PAH fluorescence is found peripheral to the ionized gas) and in the radio. However, no such association between MSX at 8.3~$\mu$m and the radio is seen in the area of non thermal emission such as for SNRs. The right panel of Fig.~\ref{fig7} shows an absence 8.3~$\mu$m emission, confirming again the non thermal nature of this remnant. As a caveat we note SNRs can be very complex in their structure and emission components where both non thermal and thermal emission from (neighbouring)  \HII\ regions can be mixed.

\begin{figure}
\center
\includegraphics[width=230pt,height=230pt, bb = 0 0 506 540 clip,]{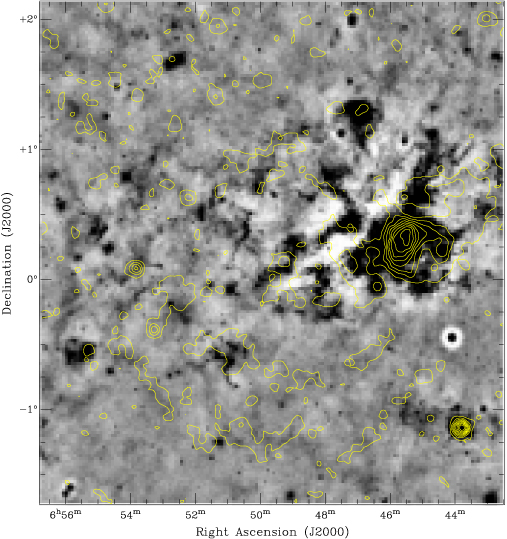}
\caption {WISE band 4 (22~$\mu$m) mosaic image of the area around G213.0-0.6 and \HII\ region S284 overlaid with PMN contours (shown in yellow). It is clear that there is no match in morphological form between the radio (and also optical  - see Fig.~\ref{fig1}) and WISE band 4 (and~3) images. Most probably the clouds seen in the  WISE image(s) belong to the S284 \HII\ region.}
\label{WISEPMN}
\end{figure}

We also compared the radio and optical emission from G213 with MIR data from the space-based Wide-field Infrared Survey Explorer (WISE).  WISE bands 1 and 2 (e.g. at 3.4 and 4.6~$\mu$m) did not show any nebular structure in this area that can be obviously correlated with G213.0-0.6. In the longer wavelength WISE bands 3 and 4 (e.g. at 12 and 22~$\mu$m) a number of fragmented dust clouds in and over the radio/optical boundaries of G213.0-0.6 are seen (see Fig.~\ref{WISEPMN}). Despite this there does not appear to be any particular spatial association of the clouds detected in these WISE bands with the observed radio/optical filamentary structure, neither could we find any match in their morphological forms. Further, keeping in mind that for the majority of SNRs any  infrared emission originates at the interaction zone of a shock with the surrounding ISM, we assume that  the 22~$\mu$m clouds seen in Fig.~\ref{WISEPMN} are connected with the \HII\ region S284. We again conclude that there is no obvious  connection of the \HII\ region and the SNR and therefore that they may be  at different distances.

As an example typical of many SNRs, we consider SNR G332.5-5.6 \citep[see][]{stu07c} where none of the four WISE bands  show any trace of a match with the radio data. As a counter example we also consider the case of SNR G11.2-0.3  where there is an excellent match seen
between the radio and WISE emissions over the whole circular shell of this remnant. These examples simply shows that  if infrared dust clouds from Fig.~\ref{WISEPMN} are in the surrounding regions (e.g. at the same distance) as SNR G213.0-0.6 a strong morphological match would exist. Additionally, support for non-interaction of these two objects is also clearly seen on Fig.~\ref{fig7} where they definitely appear as separate objects.

In conclusion, all the currently available multi-wavelength evidence confirms the presence of an SNR. This includes the negative radio spectral index typical for shell SNRs and  the matching \HA\ emission and filamentary structures which are also typical of optically detected SNRS and which closely follow the outlines of the radio shell. The follow-up optical spectra also reveals the shock excitation  expected of SNRs. Furthermore, the new high resolution optical imaging and relatively high resolution radio observations from the PMN survey at 4850~MHz and NVSS at 1400~MHz  all support the inference that the  supernova explosion occurred at a position closer to G213.3-0.4  which corresponds  to the geometric centre of the main body of the shell seen in \HA\ and of the X-ray source. This  could not be accurately estimated from the previous low resolution radio observations of \citet{Bons79} and \citet{Rei03}. We therefore re-designate the SNR G213.0-0.6  as G213.3-0.4. Finally, the \HII\ S284 region is shown to have no clear connection to the SNR contrary to what has been previously implied and so we cannot confidently use any distance determination to this \HII\ region to infer a physical size for the remnant.

\section{Acknowledgements}

We are thankful to the South African Astronomical Observatory Time
Allocation Committees for enabling the spectroscopic follow-up to
be obtained. We  thank to Kyle de Pew for assistance during these observations. QAP thanks the ANSTO access to major facilities fund for providing support for the observing run.

\end{document}